# Energy storage in structural composites by introducing CNT fiber/polymer electrolyte interleaves.

*Evgeny Senokos[1,2,3], Yunfu Ou[1,2], Juan Jose Torres[1], Federico Sket[1], Carlos González[1,2], Rebeca Marcilla[3], Juan J. Vilatela[1]\**

[1] IMDEA Materials Institute, c/ Eric Kandel 2, Getafe 28906, Madrid, Spain

[2] E. T. S. de Ingenieros de Caminos, Universidad Politécnica de Madrid, 28040 Madrid, Spain

[3] IMDEA Energy Institute, Parque Tecnológico de Móstoles, Avda. De la Sagra 3, 28935 Móstoles, Madrid, Spain

ABSTRACT

This work presents a method to produce structural composites capable of energy storage. They are produced by integrating thin sandwich structures of CNT fiber veils and an ionic liquid-based polymer electrolyte between carbon fiber plies, followed by infusion and curing of an epoxy resin. The resulting structure behaves simultaneously as an electric double-layer capacitor and a structural composite, with flexural modulus of 60 GPa and flexural strength of 153 MPa, combined with 88 mF/g of specific capacitance and the highest power (30 W/kg) and energy (37.5 mWh/kg) densities reported so far for structural supercapacitors. In-situ electrochemical measurements during 4-point bending show that electrochemical performance is retained up to fracture, with minor changes in equivalent series resistance for interleaves under compressive stress. *En route* to improving interlaminar properties we produce grid-shaped interleaves that enable mechanical interconnection of plies by the stiff epoxy. Synchrotron 3D X-ray tomography analysis of the resulting hierarchical structure confirms the formation of interlaminar epoxy joints. The manuscript discusses encapsulation role of epoxy, demonstrated by charge-discharge measurements of composites immersed in water, a deleterious agent for ionic liquids. Finally, we show different architectures free of current collector and electrical insulators, in which both CNT fiber and CF act as active electrodes.

INTRODUCTION

The rapid development of mobile electric technologies such as portable electronics, electric vehicles, vessels and aircraft has created considerable demand for energy storage systems with higher gravimetric and volumetric efficiency. The traditional approach to minimize the mass of energy storage components consists in increasing energy and power densities with respect to state-of-the-art materials. Another approach involves developing multifunctional materials combining, for example, structural and electrochemical energy storage functions, and providing a weight/volume reduction compared with mono-functional systems.

The concept of structural energy storage has been explored in batteries[1–4], supercapacitors[5–9], dielectric capacitors[10–12] and fuel cells[13,14]. Amongst these, structural supercapacitors are particularly attractive because of their relatively simple structure. Energy storage in supercapacitors is based on electrostatic charge accumulation at the electrode/electrolyte interface, typically realized in a sandwich structure of two carbon porous electrodes separated by a membrane and embedded in, or infiltrated by a liquid electrolyte with high ionic conductivity. Conventional supercapacitors provide relatively high energy and power densities, long cycle life, good reversibility and operation in a wide range of temperatures.

The development of structural supercapacitors and batteries has been reviewed recently.[15–17] Most work so far has focused on using carbon fibers (CF) as active material and a solid polymer electrolyte as structural matrix. Snyder et al., for example, examined chemical activation and coating with a redox active polymer as method to increase the inherently low capacitance of CF fabric.[6] The structural polymer electrolyte consisted of a cross-linked network of tetraethylene glycol dimethacrylate and methoxy-poly(ethylene glycol)550-acrylate with lithium imides dissolved in it. These strategies lead to samples with specific capacitance of 35 mF $g^{-1}$, tensile modulus of 10 GPa and lap shear strength of 0.75 MPa. Qian et al. developed a method to grow a high specific surface area (SSA) carbon aerogel (CAG) around CF fabrics by infusion of a resorcinol formaldehyde resin and subsequent carbonisation.[7] Combined with a poly(ethylene glycol) diglycidyl ether (PEGDGE) matrix containing 10% IL, it led to calculated power and energy densities of 0.60 F $g^{-1}$, 0.033 W $kg^{-1}$ and 0.84 mWh $kg^{-1}$, respectively. Very importantly, the presence of the CAG improved in-plane properties and produced composites with 895 MPa shear modulus and 8.71 MPa shear strength. Other strategies explored to increase the SSA of CF include the integration of activated carbon[5] or carbon nanotubes (CNT)[18]. Overall, the predominant approach has been to modify a CF laminate structure to enable energy storage. While this has led to impressive mechanical properties, similar to traditional structural composites, electrochemical properties have been difficult to realize. Often, the electrochemical characterization of CF-based structural supercapacitors requires unconventional charge-discharge measurements at low voltages 0.1V and extremely low current densities (0.001-0.1 mA cm$^{-2}$).[5,18]

An alternative approach is to build a structural supercapacitor around unidirectional fabrics of CNTs. The fabrics are arrays of macroscopic fibers of CNTs, which combine filament mechanical properties in the high performance range, with a large SSA of 256 m$^2$ g$^{-1}$ and electrical conductivity 3.5×10$^5$ S m$^{-1}$ when assembled as fabrics[19]. These materials can thus function as active material and current collector.[20]

Recently we demonstrated a simple stamping route to produce large-area all-solid supercapacitors combining CNT fibers with a polymer electrolyte containing 1-butyl-1-methylpyrrolidinium bis(trifluoromethanesulfonyl)imide (PYR$_{14}$TFSI) ionic liquid and a conventional poly(vinylidene fluoride-co-hexafluoropropylene) (PVDF-HFP) thermoplastic. The devices have high electrode-normalised values of capacitance (28 F g$^{-1}$), energy (11.4 Wh kg$^{-1}$) and power (46 kW kg$^{-1}$) densities, together with exceptional flexibility and a specific tensile strength (577 MPa SG$^{-1}$) above that of copper or a high-performance polymer[21]. Improvements in the collection of fabrics are expected to produce devices with strength of 0.37 GPa SG$^{-1}$ and modulus 18 GPa SG$^{-1}$ by realizing tensile properties of individual CNT fiber filament and reducing thickness of the polymer electrolyte membrane. Similarly, gravimetric energy density of CNT fabrics has been recently shown to increase by coating with a pseudocapacitive metal oxide deposition[22] or by gas-phase functionalization of the material[23]. This combination of properties makes CNT fiber-based electrodes an ideal element in structural supercapacitors.

In the present work we produce a new type of energy storing structural composite by embedding all-solid thin electric-double layer supercapacitors (EDLC) as interleaves between plies of CF, then infusing with a thermosetting polymer and curing. We present a thorough study of electrochemical properties in-situ during composite fabrication by vacuum bag resin infusion, as well as during bending tests. Such measurements demonstrate that the envelope of electrochemical operation is much wider than the mechanical deformation expected in a laminate composite. The samples in this study combine high mechanical properties and superb energy storage capabilities with the highest values of energy and power densities reported so far for structural SC devices. The work also explores methods to improve interlaminar properties by using grid-shaped interleaves that thus mechanically interconnect CF plies, as confirmed by synchrotron 3D X-ray tomography. Finally, we present different architecture to electrochemically interconnect the CNT fiber electrodes and CF layers.

RESULTS AND DISCUSSION

**Energy storing composite fabrication and in situ electrochemical characterization.**

Figure 1a depicts the fabrication process of the structural EDLC composites. Overall, the method consists in producing a lay-up of CF (Hexcel G0926) with a thin EDLC interleaf between layers, and then infusing the fabrics with a thermosetting resin (Ashland Inc. Derakane 8084). The interleaf is itself a thin sandwich structure comprising a polymer electrolyte membrane (100-120 µm) between two CNT fiber sheets deposited on a thin aluminum support. Such sandwich structure is produced by simply applying a small pressure to achieve impregnation of the soft membrane into CNT fiber porous electrodes.[21] The self-standing EDLC can be large (100 $cm^2$), although the samples required for this study were smaller, around 4 $cm^2$, dictated by the size of a typical composite beam subjected to bending tests. Interleaves were laid down between CF fabric layers, in most cases using insulating plastic tape to avoid the CF short circuiting the device.

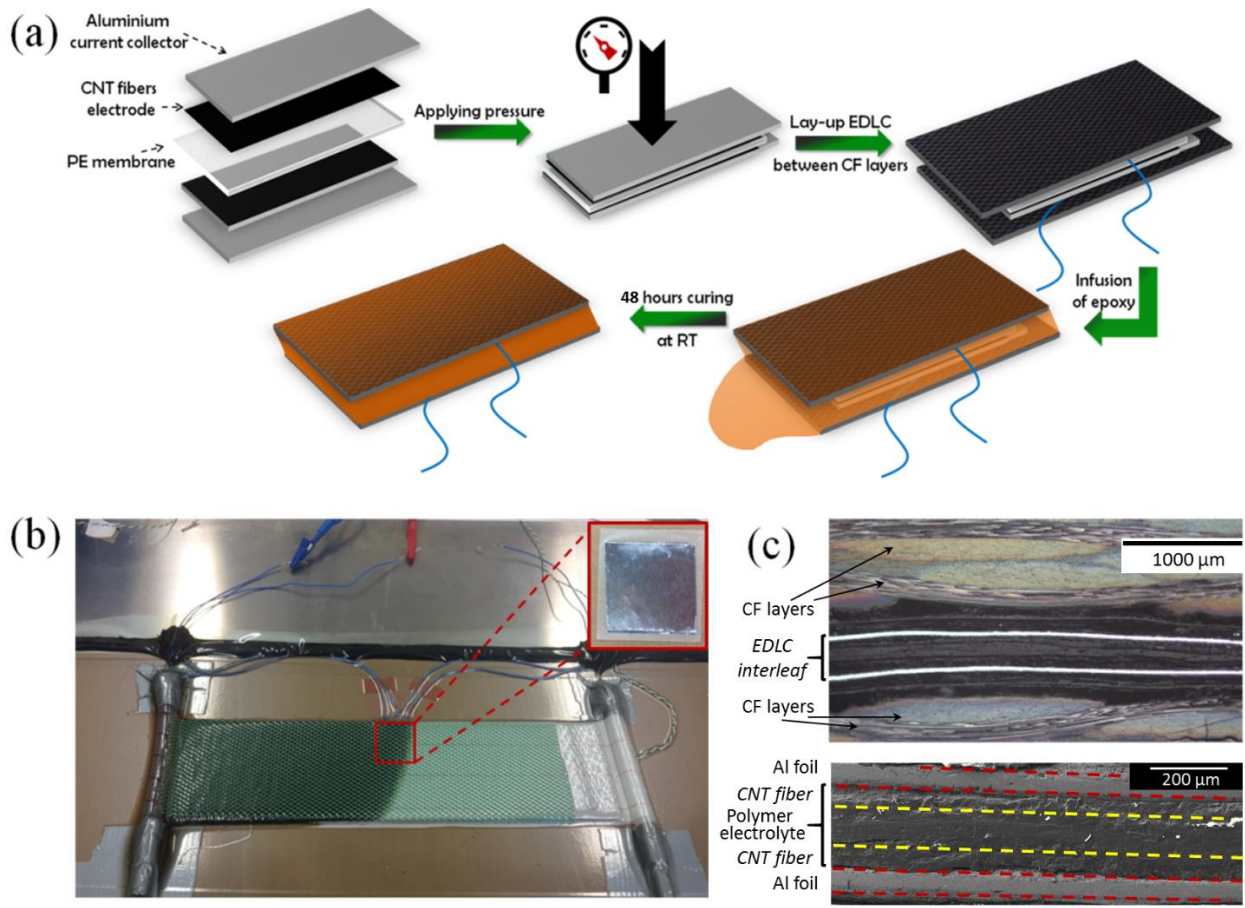

**Figure 1.** Infusion of epoxy into CF/EDLC/CF lay-up. (a) Scheme of the fabrication of structural supercapacitor composite by stamping a CNT fiber-based EDLC interleaf, embedding it between CF plies and infusion/curing of epoxy resin. (b) Photographs of a CF/EDLC/CF lay-up during epoxy infusion and of the 4 cm$^2$ EDLC interleaf embedded in it (inset). (c) Optical micrograph of composite cross-section (top) showing successful integration of EDLC/CF/epoxy in the laminate and scanning electron micrograph (bottom) of integrated EDLC interleaf.

Composites were then fabricated by vacuum bag resin infusion using a [0º]$_8$ lay-up. Briefly, this consists in producing resin flow along the fabric lay-up driven by a negative pressure gradient. Figure 1b shows a photograph of the composite during fabrication, with the cables attached to the device visible. The procedure of structural SC fabrication achieves good integration of the components in the CF/EDLC/ epoxy laminate. Figure 1c shows optical and electron micrographs of the laminate cross-section showing the EDLC interleaf inbuilt between CF layers in the cured sample. The resin was able to fill the carbon fabric with the EDLC device, allowing formation of a continuous ply structure without large voids and defects. Synchrotron tomography measurements confirm that the insertion of the EDLC interleaf does not impede adequate flow of the resin surrounding the interleaf and the subsequent consolidation of the CF fabric layers. Small voids are detected only at the EDLC interleaf edges (Figure 1S, Supplementary Information). Tensile modulus was similar for samples with (64 GPa) and without (62 GPa) interleaves.

In situ charge-discharge tests were performed during infusion, gelation and further curing of epoxy resin in order to evaluate the impact of these processes on electrochemical performance of the supercapacitor. CD profiles in Figure 2a shows negligible changes in electrochemical behavior of EDLC after infusion. Coulombic efficiency remains high, at 99%, and equivalent series resistance (ESR) low, at 24 Ohm cm$^2$. After curing of epoxy, which takes 48 hours at room temperature, there is a small ESR increase to 35 Ohm cm$^2$, which we attribute to local damage to the sample after removal the peel-ply and distribution media layers used for fabrication of the composites. Nevertheless, as Figure 2b shows, energy density remains fairly constant at 6.40 Wh kg$^{-1}$ (normalized by mass of active material) during epoxy infusion and after curing. Inspection of devices confirms that epoxy does not penetrate into the EDLC interleaf. The high energy density of the sample under full EDLC operational conditions (3.5V and 20 mA cm$^{-2}$) is possible because of the thin layer multifunctional architecture used. Previous capacitive structural composites have typically suffered from a very high ESR (0.42-300 kOhm cm$^2$) and hence only been tested at 0.4-40 μA cm$^{-2}$.[5,7,27]

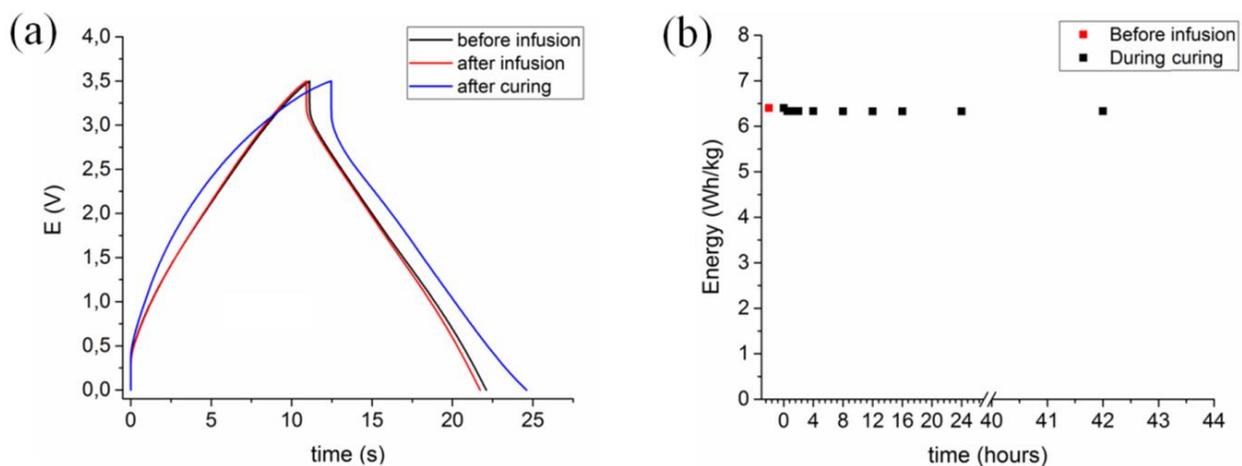

**Figure 2.** Electrochemical properties of structural EDLC composite during fabrication by vacuum bag resin infusion. (a) Charge-discharge profiles at 5 mA cm$^{-2}$ of EDLC before infusion, after infusion and after curing of epoxy resin. (b) Energy density obtained for different stages of infusion and curing.

**In situ electrochemical characterization of structural EDLC during mechanical tests.**

The high efficiency of the structural EDLC composite shown above enables the simultaneous study of electrochemical and mechanical properties. For this work, we have chosen to carry out in situ charge-discharge and impedance spectroscopy measurements during 4-point bending on rectangular cross-section beams according to ASTM D7262 standard (Figure 3a).[26] Maximum flexural stress σ and strain ε in the beams were calculated from the total force recorded with the load cell and the displacement of the actuator of the testing frame. In order to study different ply level stress states, we produced a composite with 8 CF layers $[0]_8$ and three 4 cm$^2$ EDLC interleaves. The EDLC interleaves were placed between the first and second layer, (SC_1T), in the neutral plane (SC_2M) and between seventh and eighth layers (SC_3B), respectively. They were tested individually after composite fabrication and showed nearly identical electrochemical performance and could, for example, power three red LEDs (Figures 2S, 3S, Supplementary Information). Their position in the laminate enables studying their properties under compression (SC_1T), neutral (SC_2M) and tension (SC_3B) stress states. Figure 3a shows photographs of the sample in the unstrained state and under a large deflection (~ 30 mm). In the elastic regime we observed no indication of mechanical reduction of bending modulus due to the presence of the 3 EDLC interleaves (Figure 3b). The chord modulus obtained from the σ-ε curve (Figure 3b) was Δσ/Δε≈52GPa, which was within the range of the control sample without interleaf. This fact indicates a good stress transmission between the plies of the CF and the EDLC interleaves during bending. Electrochemical measurements were performed during 4-point bending up to a bending stress of σ≈70 MPa. CD and impedance spectroscopy measurements were collected roughly every 10MPa of bending stress for each of the EDLC interleaves (Figure 4S, Supplementary Information).

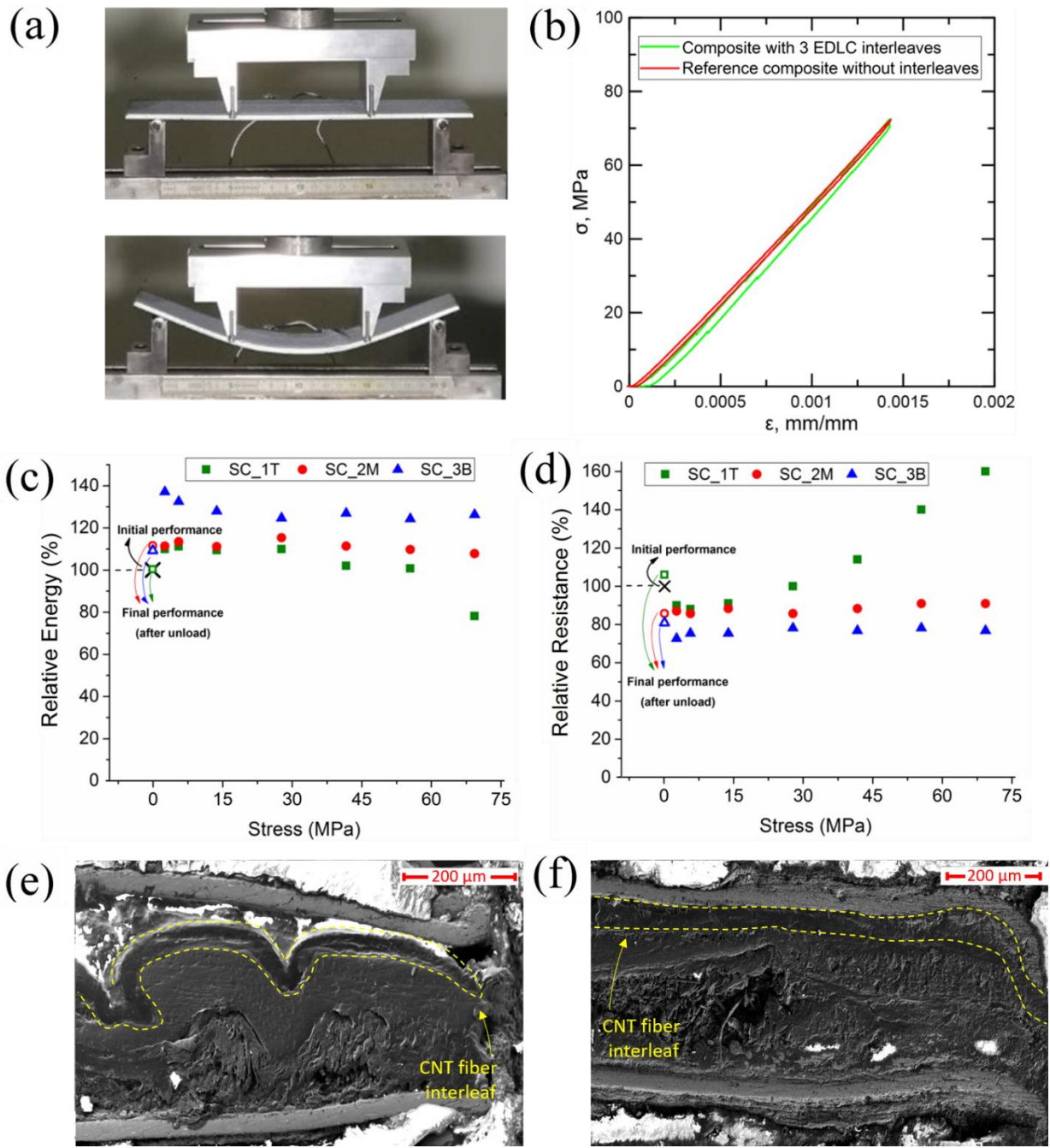

**Figure 3.** *In situ* electrochemical characterization during four point bending flexural test of the structural supercapacitor composite. (a) Photographs of the flexural test setup at the initial and bent states. (b) Comparison of stress-strain curves obtained for the structural composite containing three embedded EDLC interleaves and a reference composite produced without interleaves. (c) Relative energy and (d) ESR obtained from CD at 5 mA cm$^{-2}$ measured during bending, for the interleaves under compression (SC_1T), neutral (SC_2M) and tensile (SC_3B) stress states. (e) and (f) SEM images of the composite cross-section after the bending test. The micrographs correspond to areas near the interleaf edge, for the samples subjected to compression (SC_T) (e) and tension (SC_B) (f) stresses.

Overall, the observed change in the electrochemical properties of the interleaves is relatively small considering the large deformation of the composite structure. Figure 3c and Figure 4Sa, show that the energy density and capacitance of the EDLC interleaf under tension (SC_3B) increase by around 30% and 12%, respectively and then remains constant throughout the test. The transverse compaction during 4-point bending test results in enhanced capacitance probably due to improved infiltration of the polymer electrolyte into the porous CNT fiber electrode, as previously observed in CNT fiber-based flexible thin EDLCs under severe bending.[21] The significant change in energy density is caused by the increase in capacitance and by a drop in the ESR of the composite (Figure 3d) which was corroborated by CD test and impedance measurements (Figure 4Sa,d, Supplementary Information). Smaller increment in energy density is also observed in the interleaves in the neutral plane and the compressive region, and thus, it is attributed to a better electrical contact in the EDLC layers, probably caused by the pressure applied during 4-point bending.

The only pronounced change of electrochemical properties is observed at high bending stresses for the interleaf under compression (SC_1T), which shows a drop in energy density of 22%. This is again due to a change in ESR, which increases by about 60% relative to the unstrained state (Figure 3d and Figure 4Sb, Supplementary Information). This increase is due to local losses of electrical contact, attributed to transverse opening deformation of the device caused by the in-plane compressive stresses at this ply. The fact that the final value of ESR is lower than in the strained states suggests that most of the contact losses are reversible. Figures 3e and 3f show electron micrographs of cross sections of interleaves in the compressive ply (buckled device) and tensile ply, respectively. They show significant deformation of the interleaf under compression, in agreement with the observed changes in electrical resistance.

**Mechanical interconnection using perforated interleaves.**

While having very high power and energy densities, the EDLC interleaf layer in the composite can act as an internal defect. The polymer electrolyte in the EDLC layer is very soft, with a tensile modulus of 14 MPa. Its correspondingly low shear modulus implies that the EDLC essentially acts as a delamination, as the resistance to the opening/sliding between adjacent layers is very small. Yet, the simple structure and fabrication method used here open the possibility to produce engineered structures that mitigate the reduction in interlaminar properties. As an example, we produced an EDLC interleaf with a grid structure, that upon fabrication lead to a composite structure with adjacent lamina mechanically joined by stiff epoxy connectors (Figure 4a, b). This joining method ensures a more efficient transmission of shear stresses through the interfaces of the beam. The sample consisted of a 32 cm$^2$ EDLC with 18 holes of 1 cm diameter (~18 cm$^2$ of effective electrode area), interleaved in a composite as described above.

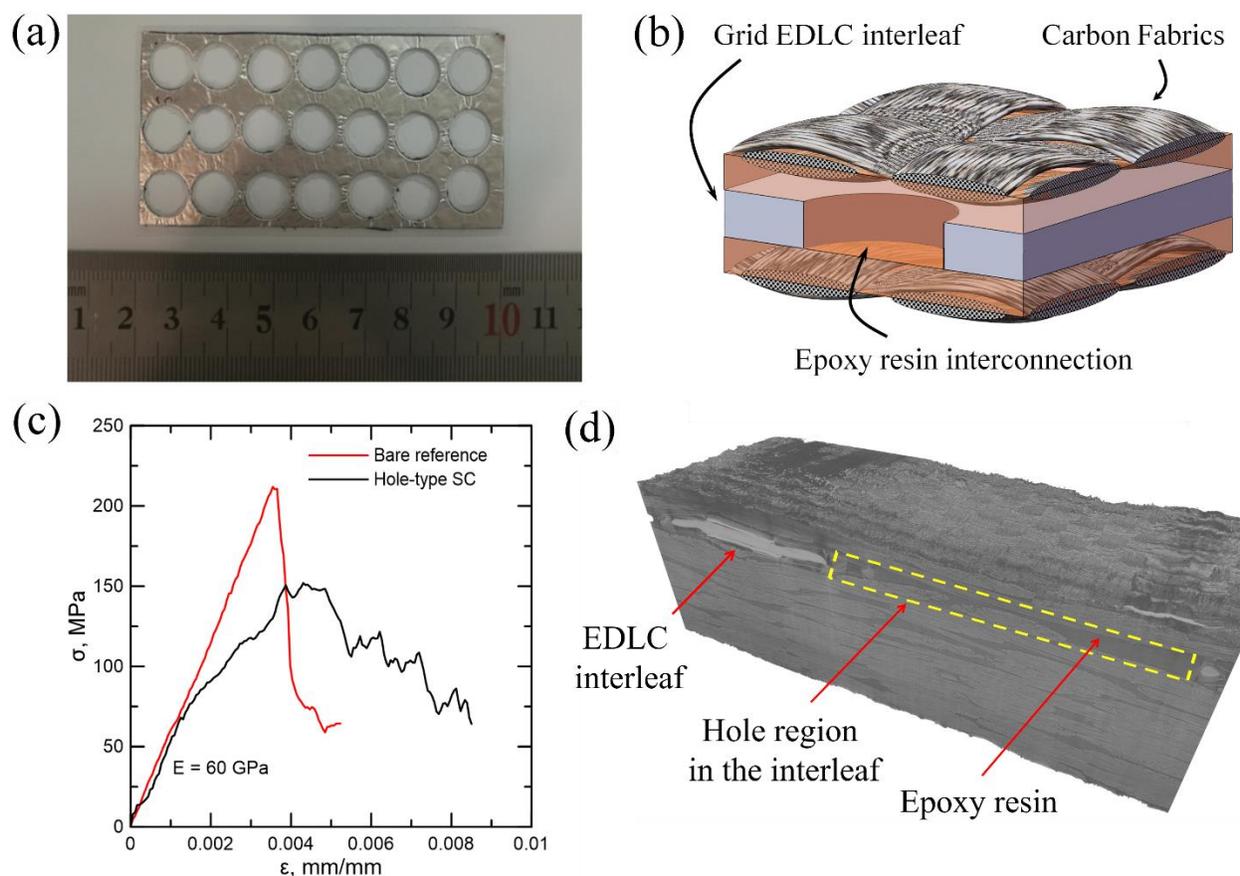

**Figure 4.** Structural supercapacitor composite with mechanically interconnected plies via the use of grid EDLC interleaves. (a) Photograph of a 32 cm$^2$ grid EDLC interleaf. (b) Schematic of the envisaged structure. (c) Bending stress-strain curves obtained for reference composite and structural supercapacitor containing a grid EDLC interleaf. (d) 3D tomography image confirming penetration of epoxy resin through channels in the interleaf.

Synchrotron X-ray tomography was used to confirm that epoxy resin successfully flowed in the through-the-thickness direction between the adjacent plies via the machined interleaf holes. Figure 4d presents a 3D image of a section of the sample after the bending test. The brighter section corresponds to the EDLC interleaf, which has higher X-ray absorbance. It is clearly delimited by a region where epoxy interconnected two adjacent plies, thus demonstrating successful production of the envisaged structure. Further characterization of this complex structure is included in SI, as color-coded 2D images extracted from different parts of the volume (Figure 5S) and as a video of the volume inspected under rotation. Inspection of the EDLC edge shows the presence of voids, which according to X-ray tomography analysis account for 1% of the volume in the full composite (SI). These voids are undesired defects, mainly caused by the presence of dielectric tapes used to insulate the electrodes electrically from the CF layer. We anticipate that avoiding these voids should be relatively simple after optimisation of the fabrication process, for example, by substituting the isolating tapes by thin polymer electrolyte membrane and using the CF as active electrode (vide infra) and evidently in composites with electrically insulating fibers.

The composite with the grid EDLC interleaf was subjected to 4-point bending tests and compared against a control sample without EDLC interleaves. Bending stress-strain curves (Figure 4c) show that both samples had a similar elastic response and bending modulus, indicating effective shear stress transmission through the CF plies. The supercapacitor composite showed evidence of non-linear deformation at around 60 MPa and failure by localized buckling at 153 MPa, whereas the control sample presented linear-elastic behavior up to fracture at 215 MPa. In the multifunctional composite failure was triggered at the interleave edges (Figure 6Sa). With the Al support and tape layers, the interleaf thickness is comparable (0.56 mm) to that of a CF lamina (0.4 mm), thus, plies are distorted in this region in order to accommodate the thickness variation imposed by the EDLC interleaf, producing stress concentrations that generate damage and local ply buckling upon loading. Current efforts are directed at removing some of the redundant elements in the interleaves, while also reducing the thickness of the polymer electrolyte membrane. Such optimized structures will lend themselves to a deeper mechanical study.

Charge-discharge profiles demonstrate nearly identical electrochemical properties for whole and perforated EDLC interleaves, proportional to the effective electrode area (Figure 7Sa, Supplementary Information). Differences observed in the Ragone plot for the two samples (Figure 7Sb, Supplementary Information) are mostly attributed to the difference in effective electrode area (32 cm$^2$ and 18 cm$^2$ for entire and perforated EDLC, respectively). Interestingly, the supercapacitor could still be charged and discharged after composite fracture, retaining its efficiency of 99%, 93% of power density and 69% of energy density (Figure 6Sb, Supplementary Information).

Considering the total mass of the 32 cm$^2$ sample, the composite has 88 mF g$^{-1}$ of capacitance, 37.5 mWh kg$^{-1}$ of energy density and 30 W kg$^{-1}$ of power density. But clearly, introducing additional interleaves between all the plies would provide a *pro rata* increase in these properties. Indeed, the highest values of energy storage obtained in this study for the composite containing three integrated EDLC interleaves are 174 mWh kg$^{-1}$ of energy density and 54 W kg$^{-1}$ of power density. Another important feature of these structures is the possibility to shift multifunctional properties between the energy and mechanical end of the spectrum, by adjusting interleaf parameters such as: the shape of the interleaves in terms of hole area, number and size of holes; and the position and number of interleaves in the laminate. Some analytical examples illustrating the envelope of multifunctional properties are included in Supplementary Information (Figure 8S, 9S).

In Table 1 we present a summary of electrochemical and mechanical performance of reported structural supercapacitors. The comparison indicates that the samples in this study have the highest values of power density (30 W kg$^{-1}$) and energy density (37.5 mWh kg$^{-1}$) reported for composites with mechanical properties in the structural range, 1 to 3 orders of magnitude above previous reports, albeit with compressive properties most certainly short of those obtained using CAGs.[7] The resulting multifunctional performance in terms of mechanical properties and particularly high energy density, is due to the extraordinary electrochemical properties of CNT fibers (high surface area and electrical conductivity), combined with the CF acting as mechanical reinforcement. The estimated value for CNT fiber-based interleaves is for non-Faradaic EDLC, but recent work on chemical functionalization[23] and deposition of metal oxides[22] of these veils suggests that further increases of 40-130% in energy density are easily within reach. Applying current models to assess performance of multifunctional structures[12,13], we observe that the introduction of higher energy-density pseudocapacitve reactions, combined with a reduction in polymer electrolyte membrane thickness, would led to a multifunctional composite structure with reduced weight relative to monofunctional systems (SI).

**Table 1.** Properties of structural supercapacitor composites

| Electrode material | Electrolyte | Cell voltage, V | Capacitance, F g$^{-1}$ | Energy, Wh kg$^{-1}$ | Power, W kg$^{-1}$ | Mechanical properties, MPa |
|---|---|---|---|---|---|---|
| CNT fiber-based interleaves in fiber reinforced polymer (This work) | Thermoplastic + [PYR$_{14}$][TFSI] | 3.5 | 0.088 | 0.037 | 30 | $s_{flex}$ = 153 $E_{flex}$ = 60 000 |
| Carbon fiber + carbon aerogel[7] | PEGDE + 10% IL | 0.1 | 0.602 | 0.001 | 0.033 | $G_{12}$ = 8710 $s_{12}$ = 895 |
| Carbon fiber[5] | PEGDE + IL + 0.1M LiTFSI | 0.1 (2, expected) | 0.052 | - (0.001, expected) | - (2.68, expected) | E = 18000 $X_C$ = 7.5 |
| CNT-grafted carbon fiber[9] | MTM57 + IL + LiFSI | 1.0 | 0.01 | 0.01 | 0.031 | $G_{12}$ = 450 $s_{12}$ = 14 E = 61200 $X_C$ = 153 |
| Carbon fiber[28] | CD552 + SR494 + 0.825M LiIm | 2.5 | 0.093 | 0.021 | 0.15 | E/SG$^a$ = 12000 $G_{12}$/SG = 310 |
| CNT fiber[21] | Thermoplastic + [PYR$_{14}$][TFSI] | 3.5 | 6.7 | 0.91 $^b$ | 3700 $^b$ | E = 790 $^c$ $s_{11}$ = 53 $^c$ |
| MWCNTs/ABA/polyaniline-modified carbon fiber[29] | CF$_3$SO$_3$Li + PEG copolymer | 1.0 | 0.125 | 0.017 | - | $s_{flex}$ = 21 $E_{flex}$ = 2900 |

$^{a)}$ Normalized by specific gravity (SG). $^{b)}$ Electrochemical properties normalized by device weight without metallic current collector or encapsulation material, $^{c)}$ Tensile test of active material and polymer electrolyte membrane.

**Additional aspects of multifunctionality.**

Apart from its mechanical functions, in energy storing composite the epoxy matrix can take the key function of encapsulation of the electrochemical system. The ingress of water in ionic liquid-based EDLCs, for example, causes a dramatic reduction of the electrochemical stability window of the ionic liquid and degradation of device properties[30], hence the use of pouch shells and other encapsulation media, which evidently increase the weight of the device. Because of its relatively low permeability to water, cross-linked epoxy is an inherent barrier for water. To demonstrate this, we tested electrochemical properties of a fractured structural EDLC composite before and 12 hours after complete immersion in water (Figure 5a). Figure 5b shows that CD profiles are nearly identical, with capacitance, power and energy density values within 4% of those in the dry state. For reference, a plain self-standing EDLC interleaf has a 45% drop in energy after 12 hours of immersion in water (Figure 10S, Supplementary Information).

Finally, we present a method to produce integrated structural composites where both the CNT fibers and the CF act as active electrodes, without recourse to metallic supports, external current collectors or plastic tape insulators. This is realized by embedding polymer electrolyte membranes between CNT fibers veils and CF tows, as schematically shown in Figure 6a. Unlike the samples discussed above, where the pre-formed EDLC was embedded between plies, in Figure 6a the EDLC composite array is assembled layer by layer. The polymer electrolyte membranes play a dual role acting both as separator, avoiding the electrical contact between CNT and CF electrodes, and as an ion conductor media connecting them ionically. As depicted in Figure 6a, electrically connected CNT fibers and CF can act as an EDLC system improving the energy storage capacities of the integrated structural composite. In fact, such CF/CNT fiber configuration was successfully charged and discharged (Figure 6b) with high coulombic efficiency (98%) and providing sufficient power to light a red LED (Figure 6c). As expected, energy and power densities of the resulting arrangement was lower than for CNT/CNT configuration (Figure 6b) due to the low surface area of the CF (0.2 $m^2g^{-1}$ vs. 256 $m^2g^{-1}$ for CNT fibers). These preliminary results open a new avenue for energy storage in multifunctional structures combining CF and CNT fibers.

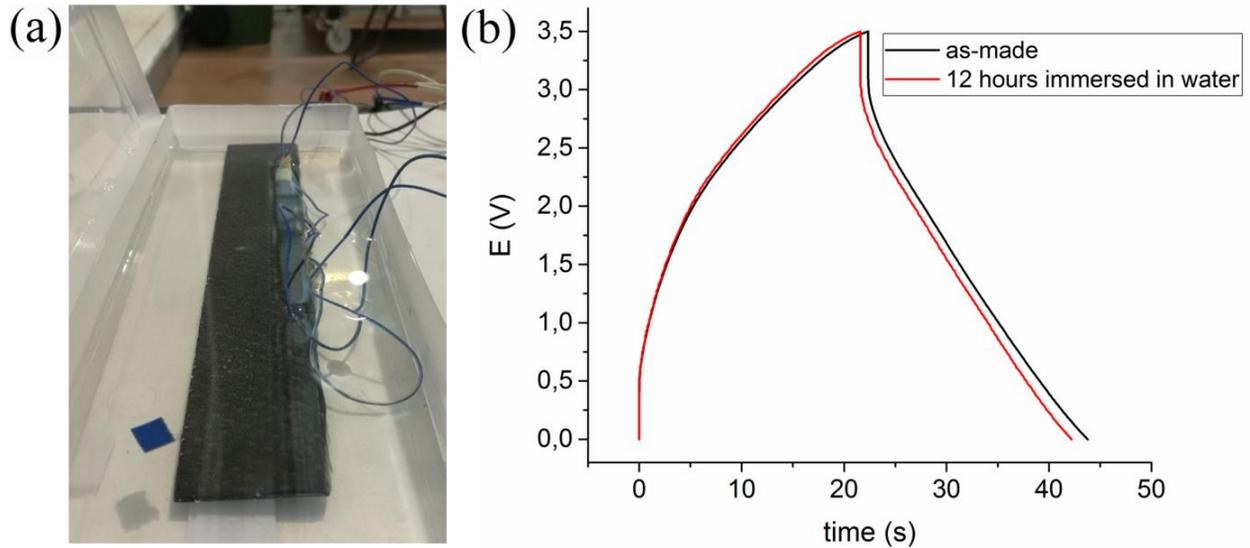

**Figure 5.** Encapsulation of the EDLC interleaf by epoxy. (a) Photo of structural supercapacitor composite immersed in water (b) Comparison of CD profiles before and 12 hours after immersion in water show nearly identical electrochemical properties.

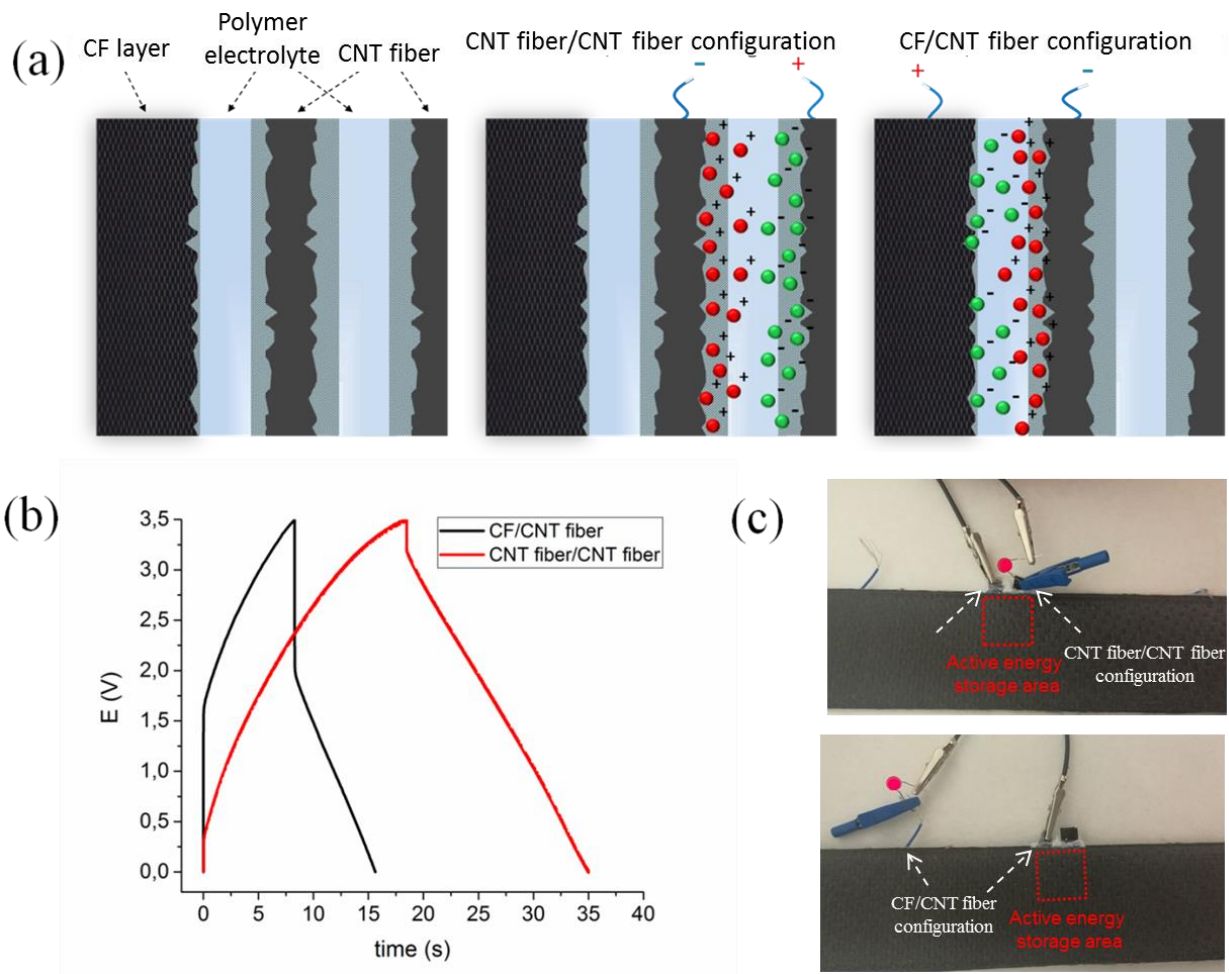

**Figure 6.** Alternative structural composite supercapacitor architectures using both CNT fibers and CF as active material, without current collector, separator or insulating layers. (a) Scheme of different EDLC configurations in the composite. (b) Charge-discharge profiles comparing CNT fiber/CNT fiber and CNT fiber/CF device configurations (c) Photographs of the structural supercapacitor powering a red LED in the two configurations.

CONCLUSIONS

In this work we present the fabrication of a novel structural composite supercapacitor based on CNT fibers/polymer electrolyte interleaves embedded between carbon fiber fabrics and infused by epoxy. The method is simple and compatible with industrial composite fabrication techniques.

Once embedded in the composite structure, the EDLC interleaves had identical electrochemical properties as before, with Coulombic efficiency of 98% and low ESR of 35 Ohm cm$^2$ even when charged at 3.5V. In situ electrochemical measurements during 4-point bending were performed on composites with EDLC interleaves at different ply positions in order to study the tensile, neutral and compressive scenarios. The results show that electrochemical properties were essentially retained under high flexural deformations and stresses (70 MPa). Changes in ESR produce 10-37% increases in energy density upon initial loading, and a drop of 22% at large strains. Even after sample failure by buckling the devices could still be charged and discharged, with 69% and 93% of the initial energy and power densities, respectively.

The composites produced here stand out for their high power density (30 W kg$^{-1}$), 1 to 3 orders of magnitude higher that the state-of-the art, combined with the highest measured energy density (37.5 mWh kg$^{-1}$) for the structural composites. Although their interlaminar properties are expected to be substantially reduced by the presence of the soft-matrix interleaves, the flexural modulus and flexural strength achieved were still 60 GPa and 153 MPa, respectively. As a demonstration of the engineering possibilities to improve interlaminar properties offered by these materials, we produced grid-shaped interleaves. Synchrotron 3D tomography shows that this design leads to epoxy regions interconnecting CF plies through holes in the interleaf. The ability to change the shape and size of the hole pattern in the interleaf open a wide range of design parameters to obtain desired composite properties. Work is in progress to remove redundant elements in the interleaves, particularly the plastic tape insulator and the Al support, and reducing their thickness. This will avoid excessive distortion of the adjacent CF and thus enable a deeper study into the mechanics of these novel structures, both experimentally and in-silico.[31] Current collector-free architectures would also enable a wider design envelope and lend themselves to compression molding, stamping, preforming and other processing steps envisaged for these multifunctional composites.

As pointers for further improvements in energy and power density, we highlight the possibility to use both CF and CNT fibers as electrodes, which reduces weight by elimination of electrical insulators while increasing the fraction of composite material used for energy storage. The energy storage properties can be also enhanced by introduction of materials producing Faradaic processes.

METHODS

**Synthesis of CNT fibers.**

Carbon nanotubes fibers were synthesized by the direct spinning method, which involves the continuous withdrawal of a CNT aerogel directly from the gas-phase during growth of CNTs by floating catalyst chemical vapor deposition,[24] using ferrocene as iron catalyst, thiophene as a sulfur catalyst promoter and butanol as carbon source. The reaction was carried out in hydrogen atmosphere at 1250 °C, using a S/C ratio to produce fibers predominantly made up of mutiwalled CNTs (MWNT) of few-layers.[25] The CNT fibers were collected directly on aluminum substrate used as suppor for EDLC assembly. The film of fibers was consolidated by densification of as-spun samples with acetone and dried at room temperature.

**Fabrication of self-standing all-solid EDLC devices.**

The fabrication of EDLC interleaves was carried out using the method recently reported by our research group.[21] Briefly, the process starts with the drying of CNT fibers at 120ºC under vacuum for 3 hours. The next step consists in sandwiching a pre-formed PE membrane between two CNT fiber electrodes of similar weight and applying a small pressure during 10 minutes with a uniaxial press (CARVER model 3853-0). The membranes have a composition of 70% of $PYR_{14}TFSI$ and 30% of PVDF-co-HFP and are fabricated by doctor blading.

**Fabrication of structural composite with embedded all-solid EDLC.**

The laminates consisted of 8 layers $[0]_8$ of Hexcel G0926 (five harness satin fabric with 370 $gm^{-2}$ of areal weight and 6K fiber yarns), with the interleaves placed between lamina and using DERAKANE 8084 elastomer-modified bisphenol-A epoxy vinyl from Ashland Inc. as infusion matrix. The structural supercapacitor devices were fabricated by vacuum assisted resin infusion. Briefly, it consists in sealing the fabric lay-up with a plastic bag and applying vacuum to infuse the fluid through the fabric as porous media. Upon opening an inlet port connected to the resin pot the resin flows through the laminate driven only by the pressure gradient between inlet and oulet. Infusion was carried out using two layers of distribution media (98 $gm^{-2}$ Airtech Green Flow 75) which ensures appropriate through-the-thickness flow from both sides of the interleaf to avoid void generation. Polyester EconoStitch 88 $gm^{-2}$ layers were used as peel-ply materials. To avoid direct contact between interlead and carbon fabrics double-sided adhesives were attached to the SC surfaces. MEKP hardener and Cobalt octoate catalyst with Derakane resin using the recommended concentration (100:1.5:0.3). The full curing reaction takes 48 hours at room temperature, with the matrix reaching its gel point after 30 – 60 min.

**X-Ray tomographic analysis of structural composites.**

Synchrotron X-ray tomography of the composite samples was performed at the TOMCAT beamline of the Swiss Light Source, Paul Scherrer Institute, Villigen, Switzerland using a monochromatic X-ray beam energy of 18 keV. X-ray images were acquired with the detector at 40 mm distance from the sample to allow for phase contrast effect. Four consecutive volumes were acquired and stitched together after reconstruction. For each individual volume, 1501 projections were measured with an exposure time of 100 ms per projection and a resolution of 1.625 µm/voxel. The detector system consisted of a 2560 x 2160 pixel CMOS camera coupled to an Optique Peter microscope system with a ×4 objective lens to magnify the visible light converted by a LuAG:Ce 20 µm scintillator. The image reconstructions were made using phase retrieval algorithm based on Paganin approach.

**Mechanical testing.**

Four-point bending on rectangular cross-section beams was carried out with a Universal electromechanical testing machine (INSTRON 3384), using a displacement rate of 1 mm/min. The tests were performed according to ASTM D7264/D7264M-07[26]. Beam dimensions were 250.0 mm x 40.0 mm x 2.9 mm, for the sample length, width and thickness, respectively. The load span and the support span set at 100 mm and 200 mm, respectively. The bending stress was calculated for every point on the load-deflection curve using the following equation:

$$\sigma = \frac{3PL}{4bh^2}, \tag{1}$$

where $\sigma$ is the bending stress at the outer surface in the load span region (constant moment), $P$ is the total applied force measured with the load cell, $L$ is support span, $b$ is width of beam and $h$ is thickness of beam. The maximum bending strain at the outer surface also occurs at mid-span, and it may be calculated as follows:

$$\varepsilon = \frac{4.36\delta h}{L^2}, \tag{2}$$

where $\varepsilon$ is maximum strain at the outer surface and $\delta$ is mid-span deflection obtained from the testing frame actuator displacement.

**Characterization of Electrochemical Properties.**

Electrochemical behavior of structural EDLCs was investigated by galvanostatic charge-discharge experiments and electrochemical impedance spectroscopy. The frequency range studied varied from 200 kHz to 10 mHz at bias voltage of 0 V with a potential amplitude of 10 mV. CD tests during infusion, curing process and bending test of structural devices was performed at 5 mA cm$^{-2}$ and 3.5 V. CD measurements at static conditions before and after curing of epoxy resin were performed at different current densities (1 – 20 mA cm$^{-2}$) and 3.5V. Capacitance of the full device ($C_{cell}$) was obtained from galvanostatic charge-discharge from the slope of the discharge curve as $C_{cell}$ = I/slope. Real energy density ($E_{real}$) and real power density ($P_{real}$) were calculated by integrating discharge curves according to the following equations:

$$E_{real} = I \int V dt \tag{3}$$

$$P_{real} = \frac{E_{real}}{t_{dis}} \tag{4}$$

ADDITIONAL INFORMATION

**Corresponding Author**

* E-mail: juanjose.vilatela@imdea.org

**Author Contributions**

E.S and Y.O fabricated samples and carried out electrochemical and mechanical testing. J. J. T. and F. S. performed tomography measurements and produced and analysed reconstructed images. C. G., R. M. and J. J. V. devised the multifunctional structures and their corresponding fabrication and test methods. The manuscript was written through contributions of all authors. All authors have given approval to the final version of the manuscript.

**Competing Financial Interest**

The author(s) declare no competing financial interests.

ACKNOWLEDGMENTS

The authors are grateful for generous financial support provided by the European Union Seventh Framework Program under grant agreements 678565 (ERC-STEM) and Clean Sky-II 738085 (SORCERER JTI-CS2-2016-CFP03-LPA-02-11), the MINECO (RyC-2014-15115, MAT2015-64167-C2-1-R) and the Comunidad de Madrid through MAD2D-CM Program (S2013/MIT-3007). Yunfu Ou appreciates the financial support from the China Scholarship Council (grant number 201606130061).